\documentclass[preprint1]{aastex631}

\usepackage{natbib}
\usepackage{xcolor}
\usepackage{amsmath}
\newcommand{\orcid}[1]{\href{https://orcid.org/#1}{\textcolor[HTML]{A6CE39}{\aiOrcid}}}
 
\newcommand\species[2]{#1 {\sc #2}}

\def\teff{\mbox{$T_{\rm eff}$}}
\def\tldr{\mbox{$T_{\rm LDR}$}}
\def\logg{\mbox{log~{\it g}}}
\def\vmicro{\mbox{$\xi_{\rm t}$}}

\def\H{\mbox{\it H}}
\def\K{\mbox{\it K}}

\shorttitle{IGRINS LDR method}
\shortauthors{Af{\c s}ar et al.}

\begin{document}

\title{EFFECTIVE TEMPERATURE ESTIMATIONS FROM LINE DEPTH RATIOS IN 
THE \H\ AND \K-BAND SPECTRA OF IGRINS}

\author[0000-0002-2516-1949]{Melike Af{\c s}ar}
\affiliation{Department of Astronomy and Space Sciences \\               
                 Ege University, 35100 Bornova, {\. I}zmir, T{\" u}rkiye}
        
\author[0000-0002-4413-4401]{Zeynep Bozkurt}
\affiliation{Department of Astronomy and Space Sciences \\                 
                 Ege University, 35100 Bornova, {\. I}zmir, T{\" u}rkiye}
                 
\author{Gamze B{\"o}cek Topcu}
\affiliation{Department of Astronomy and Space Sciences \\                
                 Ege University, 35100 Bornova, {\. I}zmir, T{\" u}rkiye}
 
\author[0000-0003-1500-332X]{Sergen {\"O}zdemir}
\affiliation{Department of Astronomy and Space Sciences \\                
                 Ege University, 35100 Bornova, {\. I}zmir, T{\" u}rkiye}
\affiliation{Nicolaus Copernicus Astronomical Center \\
                 Polish Academy of Sciences, ul. Bartycka 18, 00-716, Warsaw, Poland}

\author[0000-0002-3456-5929]{Christopher Sneden}
\affiliation{Department of Astronomy and McDonald Observatory \\           
                 The University of Texas, Austin, TX 78712, USA}
                 
\author[0000-0001-7875-6391]{Gregory N. Mace}
\affiliation{Department of Astronomy and McDonald Observatory \\           
                 The University of Texas, Austin, TX 78712, USA}
                         
\author[0000-0003-3577-3540]{Daniel T. Jaffe}
\affiliation{Department of Astronomy and McDonald Observatory \\           
                 The University of Texas, Austin, TX 78712, USA}
        
\author[0000-0002-7795-0018]{Ricardo L\'opez-Valdivia}
\affiliation{Universidad Nacional Aut\'onoma de M\'exico \\ 
                Instituto de Astronom\'ia, AP 106,  Ensenada 22800, BC, M\'exico}

\correspondingauthor{Melike Af{\c s}ar}
\email{melike.afsar@ege.edu.tr}

\begin{abstract}

Determining accurate effective temperatures of stars buried  
in the dust-obscured Galactic regions is extremely difficult from photometry.
Fortunately, high-resolution infrared spectroscopy 
is a powerful tool for determining the temperatures of stars with
no dependence on interstellar extinction.
It has long been known that the depth ratios of temperature-sensitive 
and relatively insensitive spectral lines are excellent temperature indices. 
In this work, we provide the first extensive line depth ratio (LDR) method 
application in the infrared region that encompasses both \H\ and \K\ bands 
(1.48 $\micron$ $-$ 2.48 $\micron$).
We applied the LDR method to high-resolution (R $\simeq$ 45,000) \H\ 
and \K-band spectra of 110 stars obtained with the Immersion Grating 
Infrared Spectrograph (IGRINS).
Our sample contained stars with 3200 $<$ \teff\ (K) $<$ 5500, 
0.20 $\leq$ log g $<$ 4.6, and $-$1.5 $<$ [M/H] $<$ 0.5.
Application of this method in the \K-band yielded 21 new 
LDR$-$\teff\ relations. 
We also report five new LDR$-$\teff\ relations found in the \H-band 
region, augmenting the relations already published by other groups.
The temperatures found from our calibrations provide reliable temperatures
within $\sim$70 K accuracy compared to spectral \teff\ values from the literature.

\end{abstract}


\section{INTRODUCTION}\label{intro}

The effective temperature (\teff) is of fundamental importance among stellar 
atmospheric parameters. Knowing the temperature 
enables assessment of the mass, age and surface gravity of stars, 
as well as estimation of their evolutionary status and detailed chemical abundances. 
There are several methods that can be applied for temperature estimation. 
If radius and absolute luminosity are known for a star then its
effective temperature can be directly calculated. 
The shortcoming of this method is that it can only be applied to 
nearby stars, and for more distant stars other methods are 
required. For example calibrated photometric data can be applied to achieve 
accurate \teff\ information (e.g. \citealt{alonso96, bessell98, ramirez05, masana06, mucciarelli21},
though reliable color excess values are needed for robust results. 

Several alternative methods that have been used for decades in stellar 
\teff\ determinations are based on high resolution spectra.
Since stellar absorption lines are not affected by 
interstellar extinction (or reddening), \teff\ values determined 
from spectroscopic methods are insensitive to photometric modelling choices,
in particular to assumed extinction/reddening parameters.
A popular technique requires the absence of any correlation between the 
abundances obtained from \species{Fe}{i} lines and their excitation potential 
values. 
Such \teff\ analyses often must take into account 
possible deviations from local thermodynamic equilibrium (LTE), 
and they depend on the use of reliable oscillator strengths and damping 
coefficients.
Balmer lines also become important \teff\ indicators for stars cooler 
than 8000 K because they are almost insensitive to surface gravity in these 
stars (\citealt{gray05, Heiter02}). 
This method involves the comparison of observed Balmer lines with the 
theoretical hydrogen absorption line profiles. 
Correlations between observed and synthetic spectra (e.g. 
\citealt{Cayrel91,Prugniel11}) can also be used for \teff\ determination.

Line-strength ratios of absorption lines have long been used for stellar
temperature estimates especially in spectral classification efforts.
\cite{gray91} first proposed quantitative application 
of line depth ratios (LDRs) for accurate \teff\ derivation from optical region spectroscopy. 
This technique relies on the central depth ratio of 
carefully selected absorption lines that have significantly different
responses to \teff.
This method is based on the fact that absorption lines 
with low and high excitation potential (hereafter, E.P.) 
levels respond differently to the temperature changes. 
Spectral lines with high E.P. are less responsive to the 
change in \teff\ compared to those with low E.P. 
Therefore, the ratio of line depths is a good temperature indicator, which
makes the LDR method an attractive option for \teff\ determination whenever it is applicable.  
Other parameters such as metallicity, surface gravity,  
micro- and macro-turbulence also affect the line strengths, but here one seeks
line pairs where temperature sensitivity dominates over other atmospheric
parameters.
Moreover, taking the ratio of line depths comes with the potential advantage
of muting dependencies on atmospheric parameters that affect the absorption 
lines in the same way. 

\cite{gray91} applied the LDR method just to one atomic line pair,
$\lambda$6251.8 and $\lambda$6252.6, for 49 dwarf stars.
Since then LDR studies have expanded the technique to include other
stellar classes and other spectral features. 
\citealt{sasselov90} compared \species{C}{i} and \species{Si}{i} lines
near 1$\mu$m in Cepheids and other cool giants.
\cite{gray94} 
increased the 6200~\AA\ region list to 10 line pairs
and investigated metallicity corrections for main sequence stars, and
later \cite{gray01} provided LDRs for red giants.
A series of studies extending LDR method in wavelength region on luminosity
class was published by 
\cite{kovtyukh00,kovtyukh03,kovtyukh06,kovtyukh07}. 
LDRs applicable to a 3200-7500~K temperature range for evolved stars
were derived by \cite{strassmeier00}.
Finally \cite{biazzo07} examined optical LDRs in detail, providing quantitative
assessments of luminosity/gravity effects and the influence on rotational
line broadening on various LDR ratios. 

However, highly reddened objects suffer from 
significant interstellar dust extinction, often
making them inaccessible in the optical region for high resolution 
spectroscopic studies. Therefore to
obtain any information from the objects that are positioned, 
for example, towards the Galactic center or bulge, the only 
option is to collect data at longer wavelengths, $\lambda$~$\gtrsim$~1$\mu$m.
High-resolution infrared spectroscopy is 
the ideal tool for accessing important stellar atmospheric
information from heavily obscured regions. 
Recent studies suggest that reliable atmospheric parameters of stars can be 
derived from high-resolution infrared spectral data without any help from 
the optical region ({\"O}zdemir et al.  and Garro et al., in preparation). 

A \teff\ determination method of increasing importance in near-IR high 
resolution spectroscopy involves LDRs.
Recently, \cite{fukue15} applied the LDR method to the \H-band 
(1.4$-$1.8 $\micron$) spectral region for the first time in order to measure 
more robust effective temperatures of highly reddened stars.
They obtained high resolution (R = 20,000) spectra
of 10 well-known G and K giants/supergiants (4000 $<$ \teff\ (K) $<$ 6000) 
with Subaru/infrared camera and spectrograph. 
They found nine LDR pairs of absorption lines and determined \teff\ values
of these stars with an accuracy of $\sim$60 K. 
Later, \cite{Taniguchi18} found 81 LDR$-$\teff\ 
relations using the $\it{YJ}$-band (0.90 $-$ 1.35 $\micron$) spectra 
(R = 28,000) of nine giant stars from G to M type (3700 $<$ \teff\ (K) 
$<$ 5400) obtained with the WINERED spectrograph attached to the 1.3m 
Araki Telescope.

Recently, \cite{Jian19} investigated the effects of metallicity and element 
abundances on line depths from 11 LDR pairs (seven from \citealt{fukue15}) 
using the \H-band spectra (R = 22,500) of thousands of giant stars from the 
Apache Point Observatory Galactic Evolution Experiment (APOGEE) database 
(\citealt{Majewski17}). 
Their incorporation of abundance-related terms to the LDR$-$\teff\ relations resulted in 
about 30$-$90 K scatter in temperature determination for stars of 
3700 $<$ \teff\ (K) $<$ 5000 and $-$0.7 $<$ [Fe/H] $<$ $+$0.4 dex.
\cite{lopez19} used the line depths of \species{Fe}{i} and \species{Al}{i} 
absorption lines and the molecular \species{OH} lines to determine the 
temperature of 254 K and M dwarfs from the high-resolution \H-band spectra 
obtained with the Immersion Grating Infrared Spectrograph 
(IGRINS; \citealt{Yuk10,Park14}).

In this work, we extend LDR studies to a wider wavelength range and 
investigate the LDR$-$\teff\ relations in both \H- and \K-band, 
using the high-resolution (R $\simeq$ 45,000) spectra of 110 stars obtained 
with the IGRINS.
To the best of our knowledge, this is the first study in the literature 
to use high resolution \K\ band spectra for LDR analyses.

\section{OBSERVATIONS AND STELLAR SAMPLE}\label{observ}

High-resolution (R $\simeq$ 45,000), high signal-to-noise 
($S/N$~$>$~100) spectral data of 110 stars were 
obtained with IGRINS on the 2.7m Harlan J. Smith Telescope (McDonald 
Observatory, Texas) and 
4.3m Lowell Discovery Telescope (Lowell Observatory, Arizona) between the 
years of 2014$-$2018. 
The data were reduced using the IGRINS pipeline package version 2.2 
\citep{lee16}.
IGRINS covers the \H\ and \K\  bands (1.48 $\micron$ $-$ 2.48 $\micron$), 
simultaneously, with a small gap of about 100 \AA\ between bands. 
Details of the data reduction process can be found in, e.g., \cite{afsar16} and 
\cite{Park18}. 

\begin{figure}
\begin{center}
\includegraphics[width=11cm]{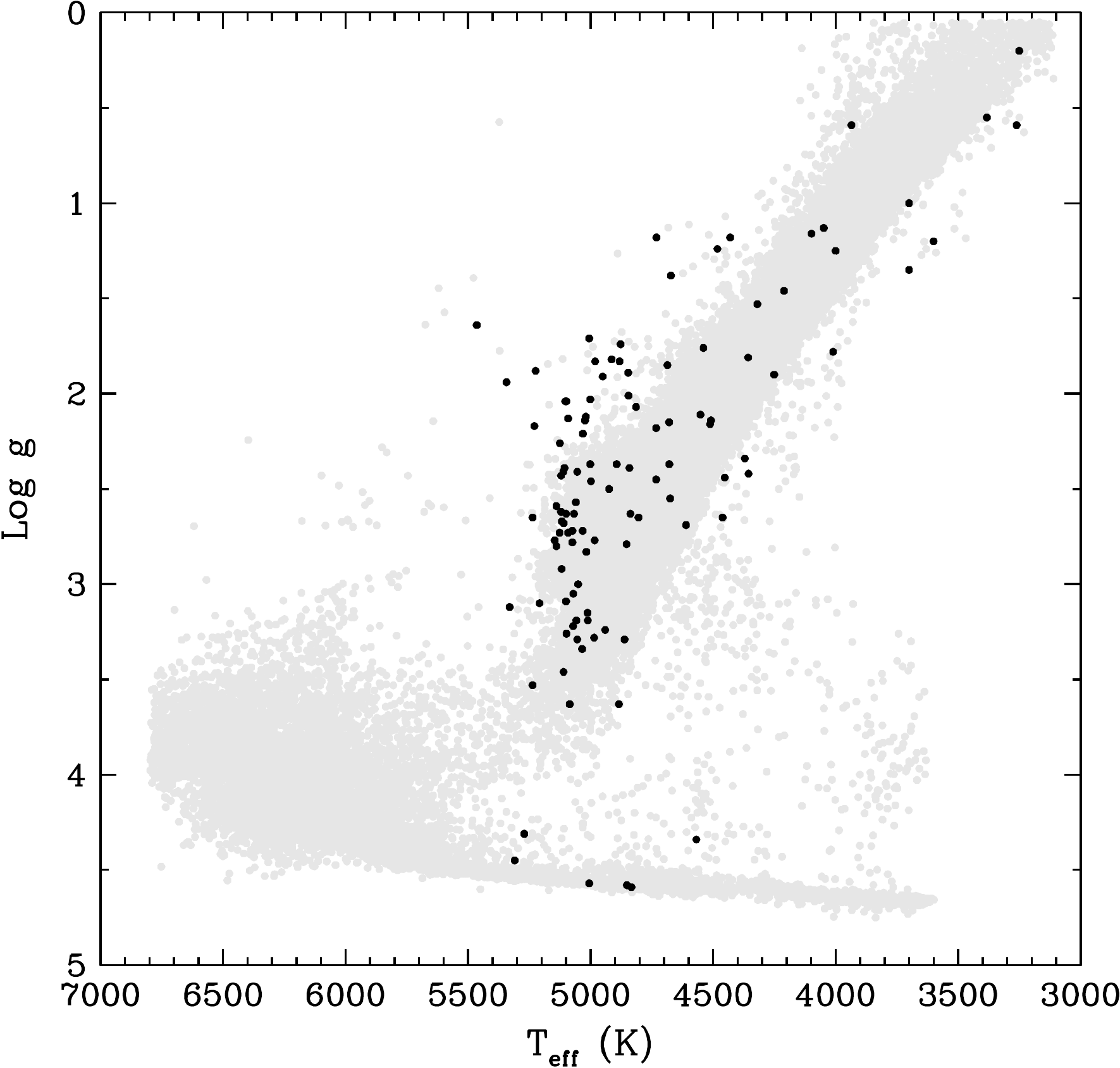}
      \caption{The mean positions of our sample (black dots) on Kiel's diagram. 
      Grey dots represent the data from APOGEE DR17 \citep{Abdurro22}.}
      \label{fig:HR}
\end{center}
\end{figure}

About 70\%\ of our sample (Table~\ref{tab-sample}) consists of red horizontal 
branch stars (RHBs, also known as secondary red clump stars) listed in 
\cite{afsar18}. 
The rest of the sample was selected from the IGRINS Spectral Library 
\citep{Park18}
mainly to expand the temperature range of our sample, but the selection of 
these low temperature stars also extended the surface gravity edges 
from about 0.2 up to 4.6. 
\textcolor{black}{ Figures~\ref{fig:HR} shows a Kiel diagram containing the positions 
of our sample (black dots) along with the stars from APOGEE DR17 (grey dots; 
\citealt{Abdurro22}). The \cite{afsar18} sample was assembled to investigate 
the properties of G-K evolved stars warmer than the red giant clump.
This accounts for the appearance of stars in Figures~\ref{fig:HR} with 
\teff~$\sim$~5000~K and \logg~$\sim$~2 that lie apart from most of the
APOGEE sample. The APOGEE DR17
data used for this diagram contains about 47000 stars with S/N $>$ 400.
The cross-match of our sample with APOGEE DR17 data yielded only 11 stars
in common. }
Overall, the stars in our study have temperature, \logg\ and 
metallicity ranges of 3200 $<$ \teff\ (K) $<$ 5500, 0.20 $\leq$ log g $<$ 4.6, 
and $-$1.5 $<$ [M/H] $<$ 0.5, as listed in Table~\ref{tab-sample}.

\section{SPECTRAL LINE IDENTIFICATION}\label{lines}

The LDR method is fundamentally based on the depth-ratio of metallic 
absorption lines with different E.P. values. As is well known
(e.g. \citealt{gray91,kovtyukh03,fukue15} and references therein), 
lines with low E.P. are more responsive to change in temperature  
than the ones with high E.P. Therefore, different line pairs with 
low- and high E.P. values are needed to establish the LDR-\teff\
relations. 
To find the useful LDR line pairs, we first carried out 
an atomic line identification process. We carefully 
examined the whole spectral region by segregating into 
wavelength intervals of 10 to 50 \AA. 
For each small interval we generated a synthetic spectrum
using the local thermodynamic equilibrium (LTE) line analysis and synthetic 
spectrum code MOOG \citep{sneden73}\footnote{Available at http://www.as.utexas.edu/$\sim$chris/moog.html}.
The process of atomic line identification was performed on the spectra 
of three stars from \cite{afsar18} that 
\textcolor{black}{ have high S/N, and}
well represent the temperature range of our entire sample: 
HIP 64378 (\teff= 4357 K), HIP 62653 (\teff= 4804 K), and HIP 54048 
(\teff= 5099 K).

During this process we made use of the following atomic line sources:
NIST \citep{Karamida21}; VALD3 \citep{Ryabchikova15}; 
ATOMIC LINE LIST v2.05b21  \citep{vanhoof18}; 
and papers based on laboratory and stellar spectra 
\citep{Shetrone15,melendez99}.
We identified the lines with the simultaneous use of synthetic  
and observed spectra displayed with the IRAF\footnote{IRAF is distributed by the National Optical Astronomy 
Observatory, which is operated by the Association of Universities for 
Research in Astronomy (AURA) under cooperative agreement with the National 
Science Foundation.} 
$splot$ task. 
Line lists needed for synthetic spectra were generated with the 
Linemake tool \citep{Placco21},\footnote{https://github.com/vmplacco/linemake} which
includes atomic lines mainly from \cite{kurucz11,kurucz18}, and also 
contains line transition data from \cite{Hartog21} (and references therein). 
This process led us to identify about 2500 atomic lines in both 
\H\ and \K\ bands.

\section{Finding LDR line-pairs and LDR$-$\teff\ relations}\label{pairs}

\begin{figure}
\begin{center}
\includegraphics[width=11cm]{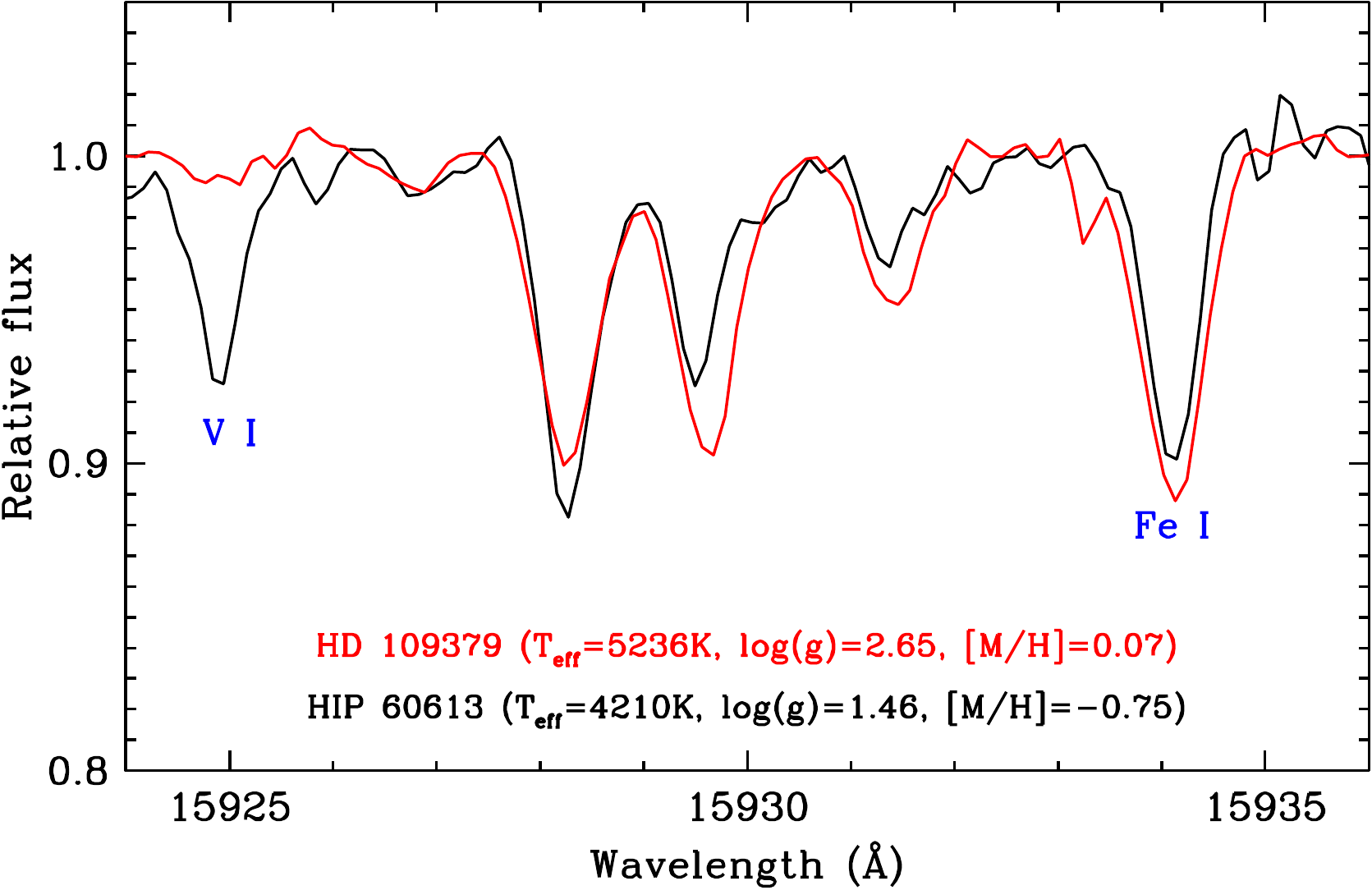}
      \caption{Temperature sensitivity of \species{V}{i} (15924.8 \AA) and 
      \species{Fe}{i} (15934.0 \AA) absorption lines in HD 109379 (\teff=5326 K)
      and HIP 60613 (\teff=4210 K).}
      \label{fig:vfe}
\end{center}
\end{figure}

After identifying atomic lines for the complete spectral region, we listed 
the elements according to their low and high E.P. values. 
After careful selection of plausible atomic absorption 
lines by eliminating other atomic and molecular contamination
as much as possible, we found about 50 potential line-pairs.
To obtain useful LDR-\teff\ relations, we first measured their central line 
depths by applying a Gaussian fitting function using the IRAF $splot$ task.
We first selected an initial set of approximately 20 stars from Table~\ref{tab-sample}
that represent the overall temperature range of our sample.
Then we measured the ratio of the line depths and plotted against spectral 
\teff\ values for this small sample.
\textcolor{black}{ 
After examining the LDR$-$\teff\ calibrations, we found 32 atomic lines
(Table~\ref{tab-lines}) to be useful for LDR method application. 
Matching these atomic lines yielded 26 LDR line-pairs 
(Table~\ref{tab-pairs}) with significant LDR$-$\teff\ relationships.}
An example line-pair of \species{V}{i} (15924.8 \AA) and \species{Fe}{i} 
(15934.0 \AA) is given in Figure~\ref{fig:vfe}.
It is important to select LDR line-pairs at wavelengths as close as possible
to minimize the errors that might originate from the continuum level setting 
(e.g. Figure 4 in \citealt{biazzo07}, and Table 5 in \citealt{Taniguchi18}). 
Unfortunately, there are not many absorption lines that can satisfy this 
constraint in the \H\ and \K-band wavelength region. 
Therefore, we were forced to pair the lines that are not in the same spectral 
order, and to use the same absorption line in different line-pairs.

\begin{figure}
\begin{center}
\includegraphics[width=12cm,angle=-90]{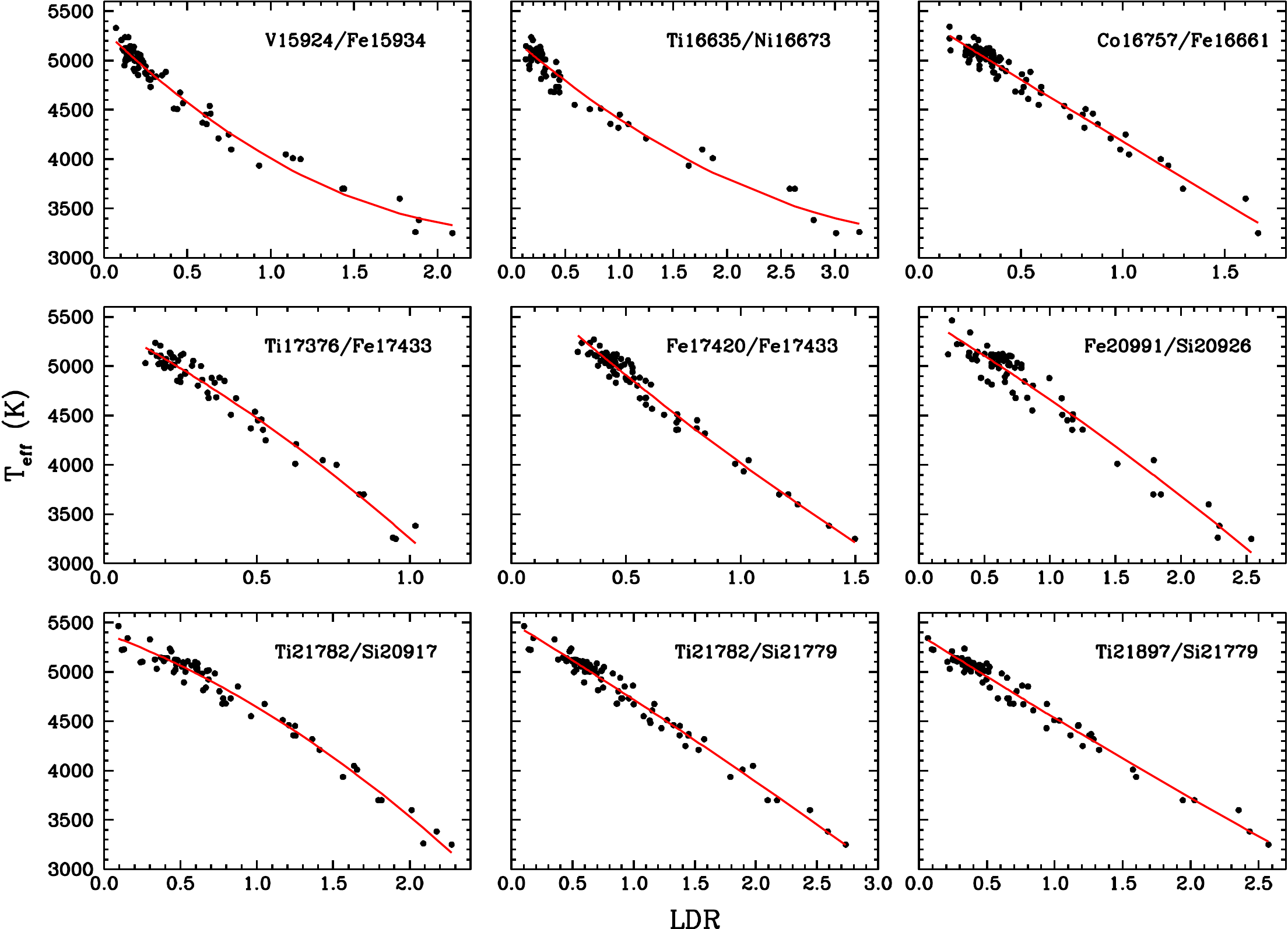}
      \caption{LDR$-$\teff\ relations. Dots indicate the stellar sample. 
      Red solid lines represent the quadratic polynomial function fitting.
      Related LDR line-pairs are given on the upper right corner in each panel.}
      \label{fig:ldr1}
      \end{center}
\end{figure}

\begin{figure}
\begin{center}
\includegraphics[width=12cm,angle=-90]{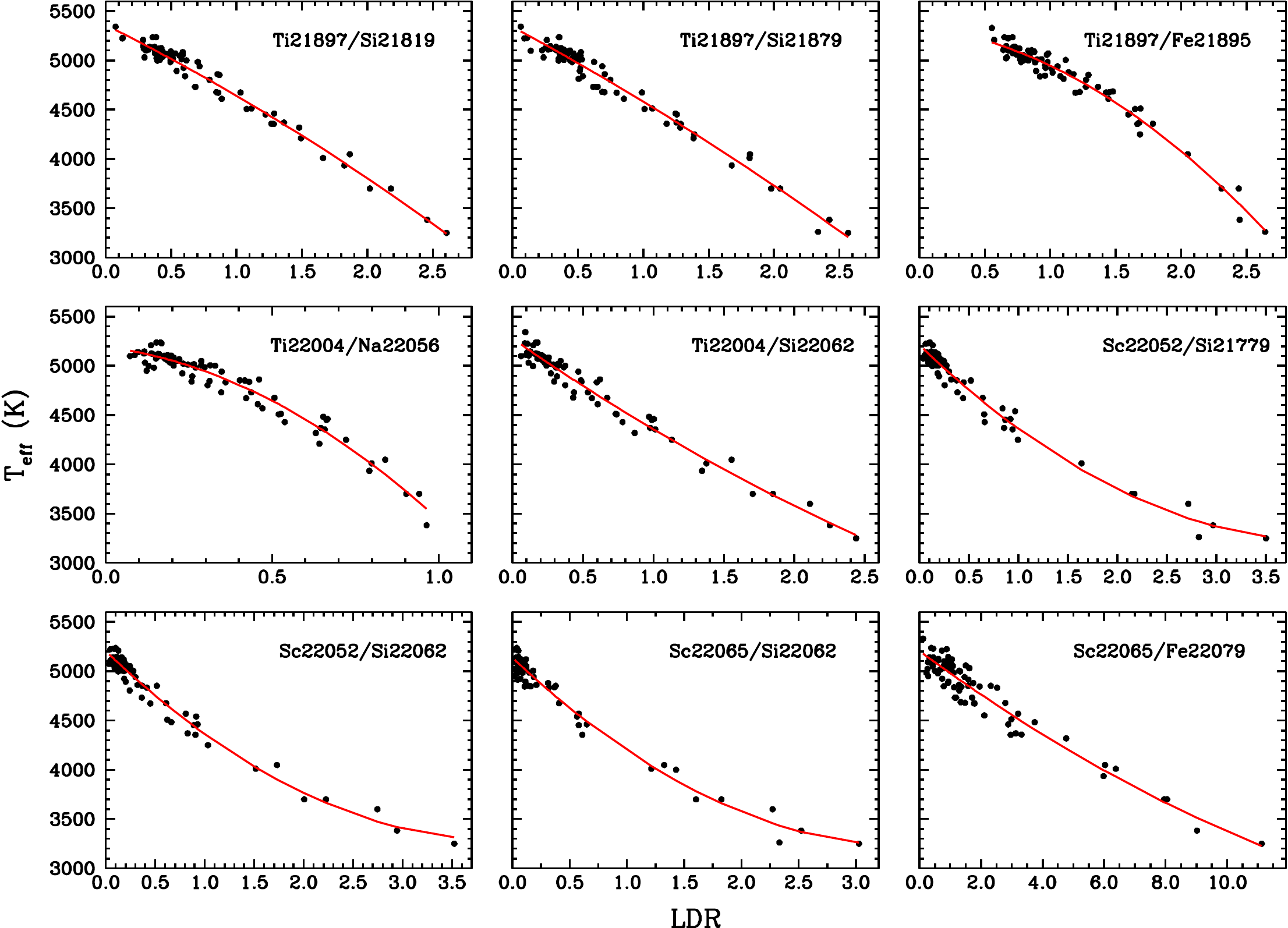}
      \caption{Same as in Figure~\ref{fig:ldr1}.}
      \label{fig:ldr2}
\end{center}
\end{figure}

\begin{figure}
\begin{center}
\includegraphics[width=12cm,angle=-90]{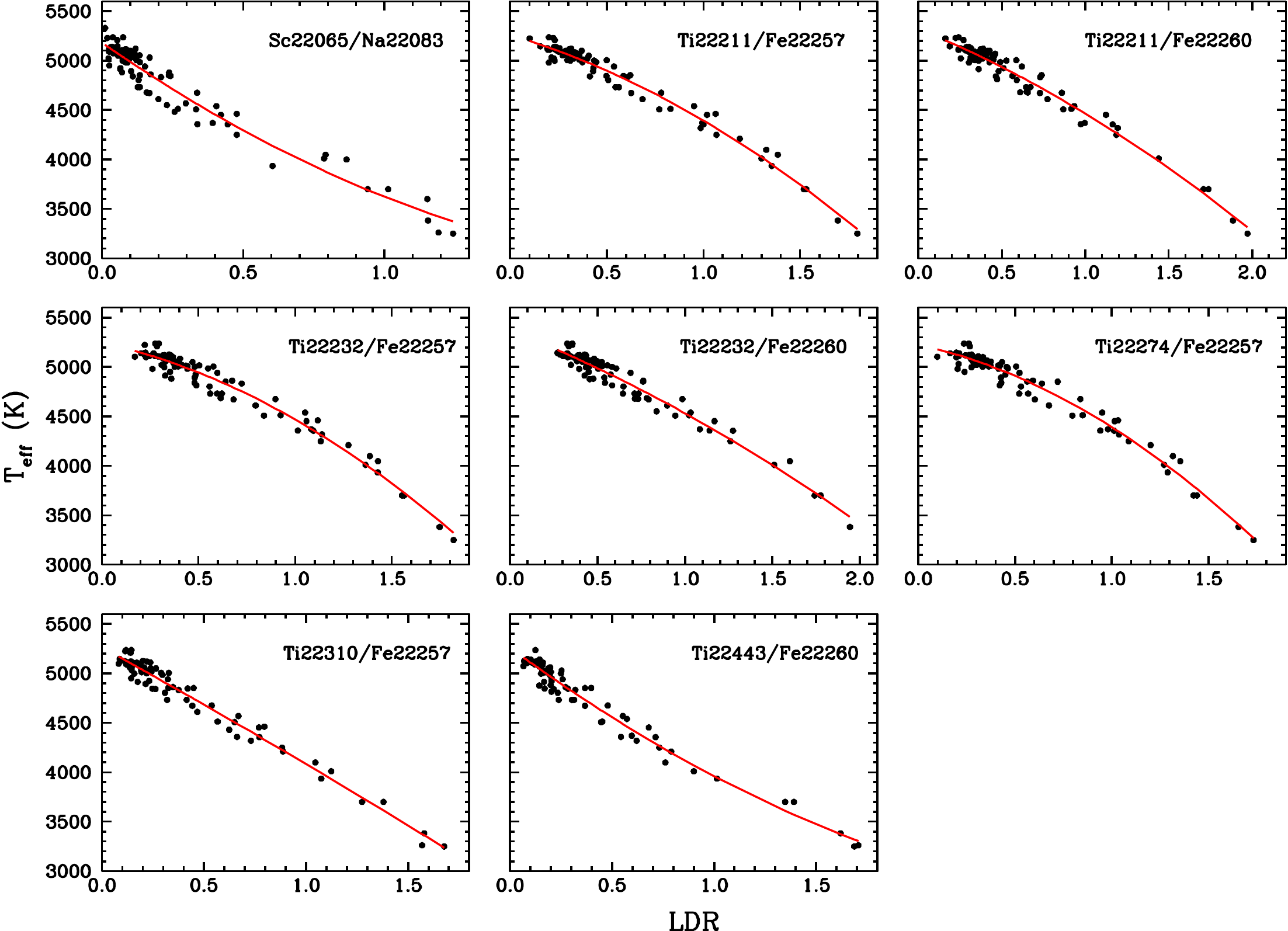}
      \caption{Same as in Figure~\ref{fig:ldr1}.}
      \label{fig:ldr3}
\end{center}
\end{figure}

Finally, we carried out the line depth measurements for the overall sample 
(Table~\ref{tab-sample}). 
\textcolor{black}{ 
The ratios of the line-pairs (low E.P. / high E.P.) were plotted against the spectral \teff\ 
values that were collected from various literature sources (the reference list for the
spectral \teff\ values is given in Table~\ref{tab-sample}).}
It should be noted here that not all the line depths could be measured for 
every star. 
For example, some weak atomic lines could not be detected in 
metal-poor stars ([M/H] $<$ $-$1). 
A couple of stars turned out to be spectroscopic binaries that had not been
previously noticed.
Additionally, some data problems such as low S/N or problematic telluric 
line removal resulted in the elimination of some of the line pairs in a
few stars.
To characterize the LDR-\teff\ relationships we used quadratic 
polynomial functions 
(\teff=a$x^{2}$+b$x$+c, where $x$ is a particular LDR) to best represent the 
distribution of data points in all of the calibrations.
To increase the sensitivity of the calibrations we ran multiple 
iterations to identify and eliminate aberrant data points. 
The method we used for iterations requires the difference between 
\textcolor{black}{ 
the actual (spectral \teff\ values, Table~\ref{tab-sample}) 
and calculated LDR temperatures ($\Delta$\teff\ = \teff$-$\tldr) to be below the
$2\sigma$ level of the $\Delta$\teff\ distribution.} We ran the iterations
until all the $\Delta$\teff\ values satisfy this requirement, 
in other words all the outlier data with values higher than 
$2\sigma$ level were removed through these iterations,
which in turn led us to use about 70 to 90 stars for the 
final LDR$-$\teff\ calibrations, depending on the line-pairs studied.
All of the final LDR$-$\teff\ relationships 
are plotted in Figures~\ref{fig:ldr1},~\ref{fig:ldr2}, and \ref{fig:ldr3}, 
and the quadratic polynomial coefficients are provided in Table~\ref{tab-pairs}.

\subsection{Uncertainties in \teff\ determination}\label{uncertain}

\textcolor{black}{ 
In Table~\ref{tab-comp}, we list the mean LDR temperatures
($\overline{\tldr}$) for the entire sample, except for HIP 79248, 
where we could not measure the LDRs from any of the line-pairs 
listed in Table~\ref{tab-pairs}.  
The $\overline{\tldr}$ temperatures were obtained by taking the simple average of
the individual LDR temperatures calculated from the quadratic polynomial 
functions given in Table~\ref{tab-pairs}.
The standard errors (SE=$\sigma$/$\sqrt{n}$) and $\sigma$ values for the LDR 
temperatures of each star range between $\pm$4 $\lesssim$ SE $\lesssim$ $\pm$60, 
and 20 $\lesssim$ $\sigma$ $\lesssim$ 150. The mean of the standard errors 
($\overline{SE}$) and standard deviations ($\overline\sigma$) 
were found as $\overline{SE}$ = $\pm$15 K and $\overline\sigma$ = 59 K, 
showing the overall internal uncertainty level of the $\overline{\tldr}$ temperatures 
obtained from each line-pair.}

\textcolor{black}{ 
We cross-matched our sample with {\it Gaia} DR3 \citep{gaia16,gaia22},
finding {\it Gaia} spectral temperatures \citep{Recio22} for 94 of our sample stars. 
We also calculated photometric temperatures for 107 stars with available
2MASS J and K magnitudes \citep{Skrutskie06} using the relations provided
by \citep{gonzalez09}.
The spectral temperatures gathered from the 
literature (\teff, see also Table~\ref{tab-sample}), \teff({\it Gaia}) spectral, 
and \teff(J$-$K) photometric temperatures are also listed in Table~\ref{tab-comp}  
for comparison. 
The uncertainty level of the temperatures resulting from the application 
of our LDR method was estimated by comparing the $\overline{\tldr}$ temperatures
with the ones gathered from the literature, \teff.
The standard deviation of the (\teff$-$$\overline{\tldr}$) differences yielded the uncertainty 
level of our method as $\sigma$(\teff$-$$\overline{\tldr}$) $\simeq$ 70 K. }

\begin{figure}
\begin{center}
\includegraphics[width=10cm]{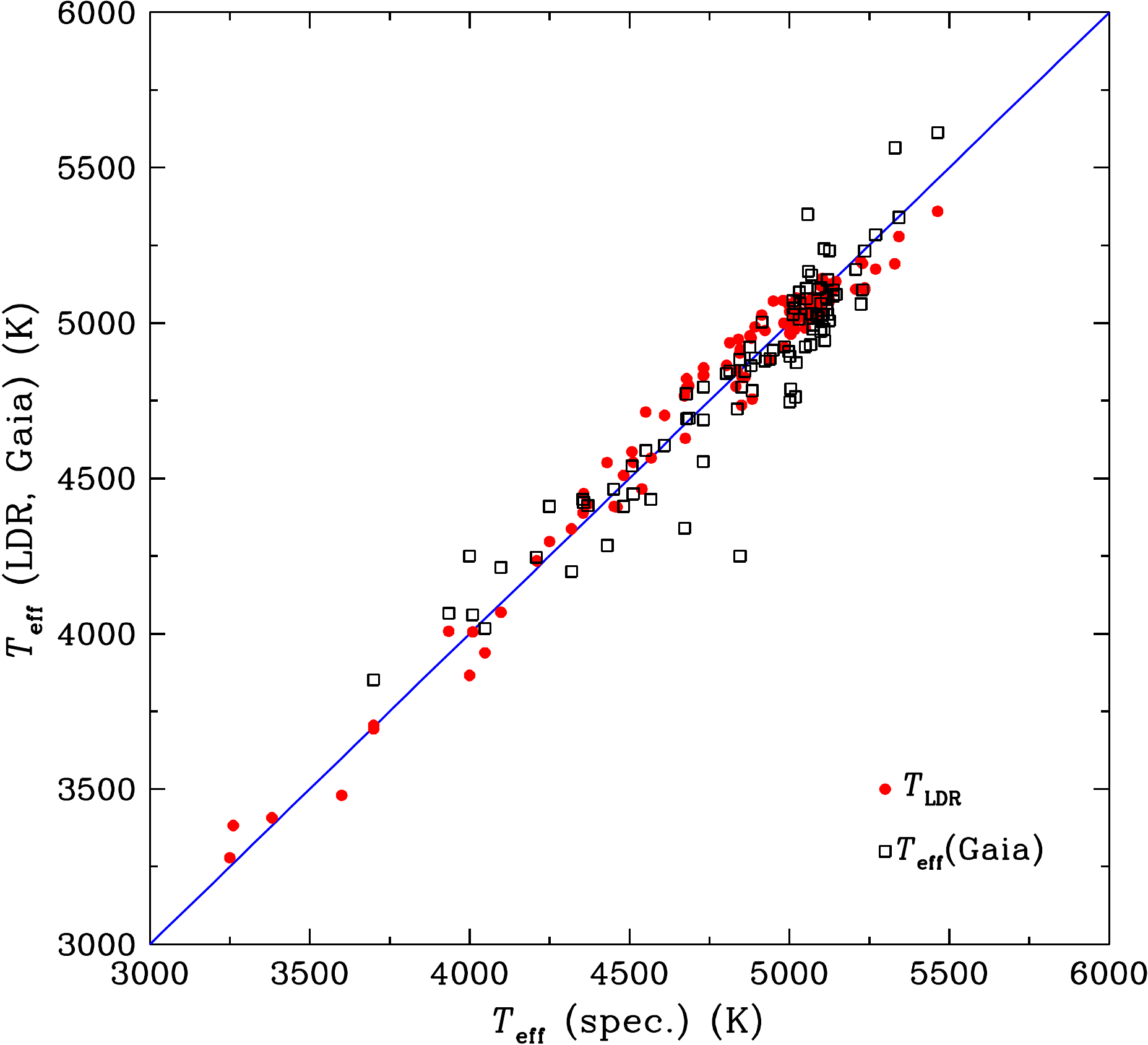}
      \caption{Comparison of the spectral \teff\ values reported in the literature
      with the $\overline{\tldr}$ (red dots) and spectral \teff({\it Gaia}) (open squares) 
      temperatures \citep{Recio22}. Blue line represents the perfect fit ($x$=$y$).}
      \label{fig:Teff_comp}
\end{center}
\end{figure}

\textcolor{black}{ 
We also compared our results with the spectral {\it Gaia} temperatures \citep{Recio22} 
and obtained $\sigma$(\teff({\it Gaia})$-$$\overline{\tldr}$) $\simeq$ 145 K. 
A plot that compares the spectral \teff\ values with the \tldr\ and \teff({\it Gaia}) 
temperatures is given in Figure~\ref{fig:Teff_comp}. The good agreement between 
the temperatures is noteworthy.}

\textcolor{black}{ Although obtained using two different measurement methods, we also compared 
our LDR temperatures with the \teff(J$-$K) photometric temperatures and found 
$\sigma$(\teff(J$-$K)$-$$\overline{\tldr}$) $\simeq$ 340 K. This substantial deviation between
the temperatures mainly originates from the J and K magnitude uncertainties reported in the 2MASS 
catalog. For example, an uncertainty about 0.25 mag in J magnitude results in an uncertainty
level of about 1000 K in the \teff(J$-$K) temperature. When we narrow down the 2MASS sample
by limiting the uncertainties in both J and K magnitudes by taking into account only the
stars with J and K magnitude uncertainties of lower than 0.2 mag, the cross-match of our
sample yields only 58 stars in common, which in turn gives 
$\sigma$(\teff(J$-$K)$-$$\overline{\tldr}$) $\simeq$ 175 K.}

\textcolor{black}{ 
Unfortunately there are not many G-K standard stars that have been observed
with IGRINS, which makes it difficult to test the accuracy level of our 
LDR$-$\teff\ relations for stars that are outside our current sample. }
The only prominent star for which we have IGRINS spectral data is Arcturus. 
Running all the LDR$-$\teff\ relations on Arcturus we obtained 
$\overline{\tldr}$ = 4263$\pm$14 K ($\sigma$ = 62 K) from 19 line-pairs, 
in excellent agreement with the temperature reported by, e.g., \cite{ramirez11} 
(\teff\ = 4286$\pm$30 K), and with the literature in general. 
The LDR$-$\teff\ relations that we did not receive reliable temperature 
information for Arcturus were mostly the ones with Sc involved (Table~\ref{tab-pairs}), 
and two additional line-pairs in the \K-band (Fe20991/Si20926, Ti21897/Fe21894)
also could not be used. In \S\ref{disc} we will discuss these line-pairs in more detail.

\textcolor{black}{ 
We also tested the LDR$-$\teff\ relations on two open cluster members;
NGC 6940 MMU 105, and NCG 752 MMU 77, of which were previously 
observed with IGRINS by our group \citep{bocek16,bocek19,bocek20}.
For NGC 6940 MMU 105 we found $\overline{\tldr}$ = 4813$\pm$10 K ($\sigma$ = 41 K) 
from 18 line-pairs, which is in good agreement with \teff = 4765$\pm$100 K from \cite{bocek16}.
The same few anomalous line pairs that we identified in Arcturus
showed a similar behaviour in MMU 105 case. 
Additionally, we could not measure the depths of two line-pairs; 
Ti17376/Fe17433 and Ti22310/Fe22257, for this star.}

\textcolor{black}{ 
For NCG 752 MMU 77, we calculated the LDR temperatures from 21 line-pairs. 
Taking an average of the individual \tldr\ values yielded 
$\overline{\tldr}$ = 4869$\pm$12 K ($\sigma$ = 60 K), which is in quite a good 
agreement with \teff = 4874$\pm$100 K from \cite{bocek15}.}
	 
Setting the continuum level in both \H\ and \K-band spectra of especially 
cool stars is difficult because they have rich spectra of many complex absorption and 
molecular features.
Despite our effort to select the absorption lines of LDR with no obvious blending issues, 
crowded spectral regions in the \H- and \K-bands are still inevitable in cool stars.
Therefore continuum placement is not perfectly defined for every LDR pair.
To be able to minimize the temperature uncertainties caused by the continuum 
placement between central line depth measurements, we preferred to place 
the continuum locally during the Gaussian line-fitting to the related 
absorption lines. 
To estimate the temperature uncertainties arising from continuum level 
placement, we used a small set of stars that well represent the temperature 
distribution of our overall sample: HD 148783, HD 119667, HIP 97599, HIP 27091, HIP 33578
(Table~\ref{tab-sample}).
We applied a simple approximation and measured the depths of the lines
six times for the same star by varying the local continuum to form a set of 30 LDR measurements 
for the individual line-pairs in both \H\ and  \K\ bands. 
The average temperature uncertainties that emerge from the LDRs obtained 
for both regions are 39$\pm$5 K ($\sigma$ = 29 K) and 23$\pm$4 K 
($\sigma$ = 22 K) for the \H\ and \K\ bands, respectively.

\section{DISCUSSION}\label{disc}

\subsection{The luminosity dependency of LDRs}\label{disc1}

The use of line depth ratios of luminosity sensitive and non-sensitive 
absorption lines for luminosity classification dates back many decades 
(e.g. \citealt{Wright65}). 
The effect of luminosity, in other words surface gravity,
on LDRs has been discussed in several studies, such as \cite{gray01}, 
\cite{biazzo07}, and \cite{Jian20}. 
The latter paper investigated the gravity effect on LDRs obtained from the 
$\it{YJ}$-band spectra of a group of stars with a \teff\ range of 3500$-$8000 K, 
including dwarfs, giants and supergiants. They found that some of the line-pairs 
they use for LDR$-$\teff\ calibrations show systematic offsets between the 
three luminosity groups. 
Their investigation led to a conclusion that the depths of some neutral lines 
(e.g. \species{Si}{i} at $\lambda\lambda$10371) in their LDR pairs are 
sensitive to \logg\, while others not (e.g. \species{Ca}{i} at $\lambda\lambda$10343).

Recently, \cite{matsunaga21} proposed a method that uses the LDRs as indicators
of the \teff\ and \logg. They found empirical relations between \teff, \logg, and LDRs, 
of which they derived using the selected line-pairs of \species{Fe}{i}$-$\species{Fe}{ii} and 
\species{Ca}{i}$-$\species{Ca}{ii} lines from the $\it{YJ}$-band spectra 
of a sample of stars from dwarfs to supergiants. 
From this method, they reported a precision level of 50 K and 0.2 dex for the 
\teff\ and \logg\ determinations, respectively.

\begin{figure}
\begin{center}
\includegraphics[width=10cm]{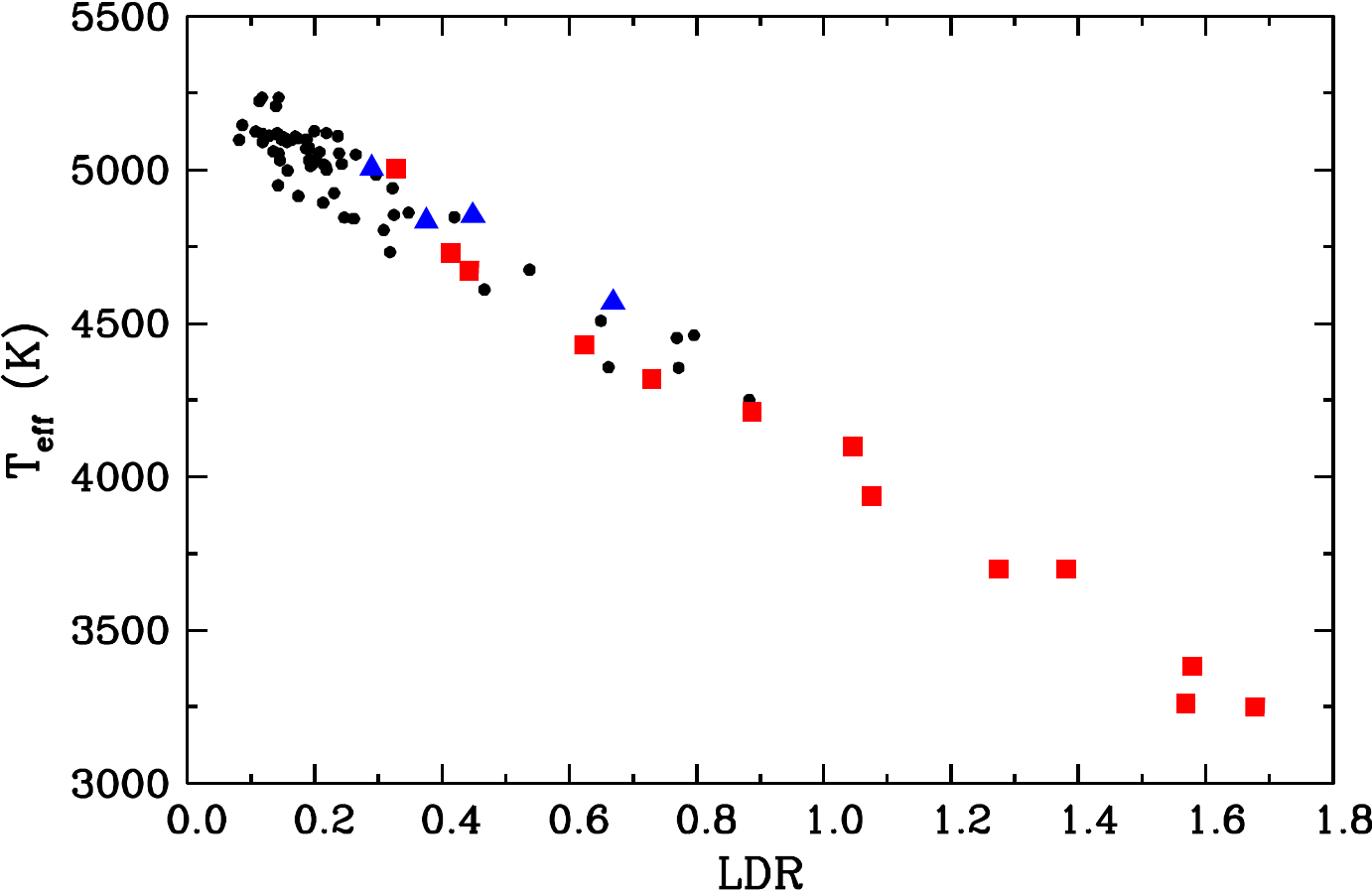}
      \caption{The LDR-\teff\ relation for Ti22310/Fe22257 line-pair. Red squares, black dots, 
      and blue triangles represent stars with 0.20 $\leq$ log g $\leq$ 1.71, 1.78 $\leq$ log g $\leq$ 3.53, 
      and 4.40 $<$ log g $<$ 4.60, respectively. }
      \label{fig:ldr_loggTiFe}
\end{center}
\end{figure}

In general, ionized species are more sensitive to the changes in gravity
than the neutral ones in cool stars,
though strong lines of neutral species such as \species{Mg}{i} b triplet 
near 5170 \AA\ (e.g. \citealt{gray05}) are known to be sensitive to the surface 
gravity changes due to mainly the pressure sensitivity of the damping constant.
Our sample has stars with surface gravities between 
0.20 $\leq$ log g $<$ 4.6, but it mainly contains
giant stars with only a few main-sequence and supergiant stars. 
The cool end of our sample contains fewer stars with \logg\ $<$ 1.75, leaving
not enough room to statistically investigate the LDR$-$\teff\ relations for 
supergiants.
On the other hand, having mostly weak lines as LDR pairs, our investigation 
of the \logg\ effect on the LDR$-$\teff\ calibrations (e.g. 
Figure~\ref{fig:ldr_loggTiFe}) showed no systematic offsets for different 
luminosity groups, allowing us to use the LDR$-$\teff\ relations as listed 
in Table~\ref{tab-pairs} without any apparent need for \logg\ correction on LDRs. 

\subsection{The metallicity effect on LDRs}\label{disc2}

\begin{figure}
\begin{center}
\includegraphics[width=14cm]{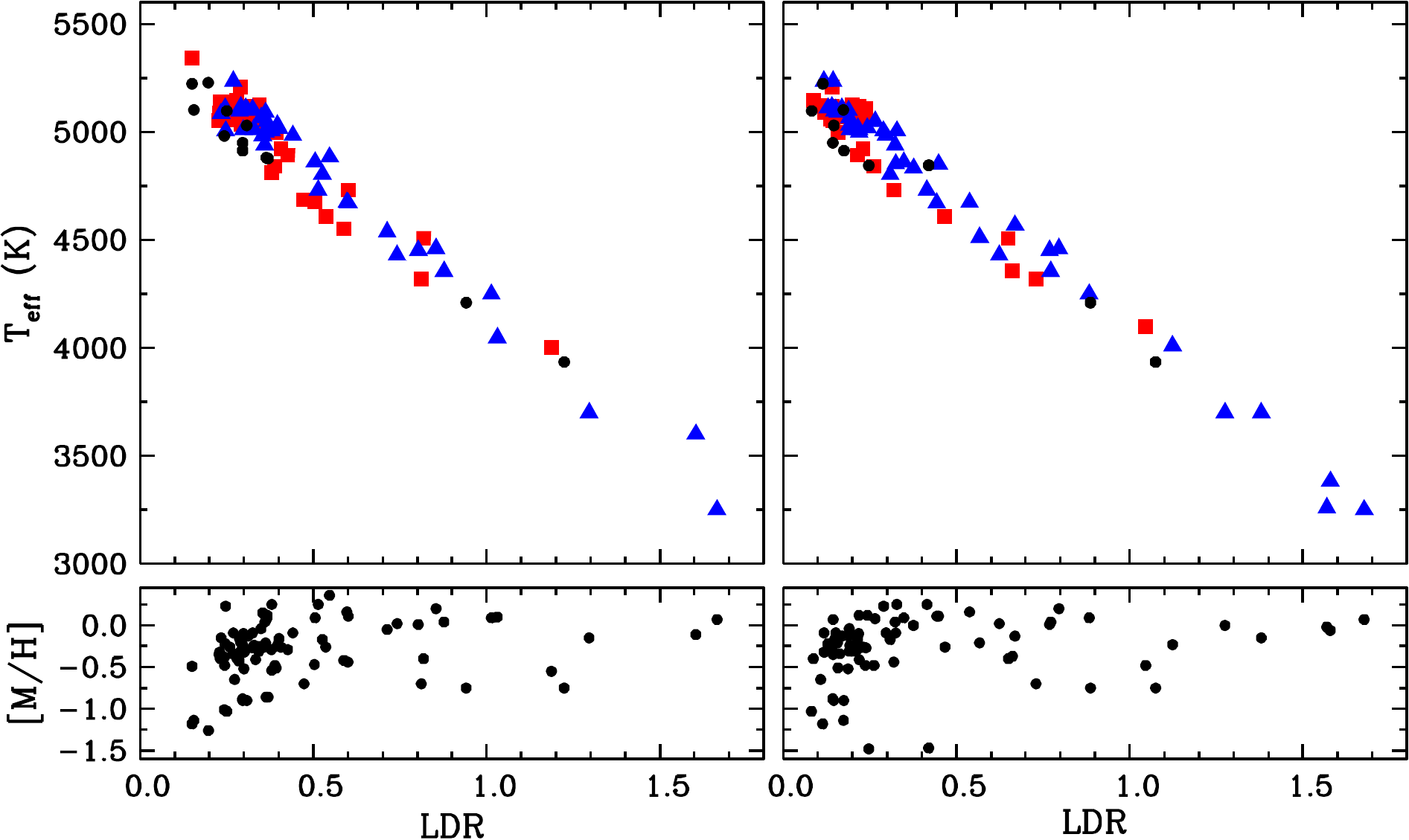}
      \caption{An example of metallicity effect on LDR$-$\teff\ relations for both \H\ and \K\ bands: 
       Co16757/Fe16661 (left panel), Ti22310/Fe22257 (right panel). In the upper panels, the stars 
       with [M/H] $<$ $-$0.75, $-$0.75 $\leq$ [M/H] $\leq$ $-$0.25, and $-$0.25 $\leq$ [M/H] $<$ 0.4
       are coloured in black (dots), red (squares) and blue (triangles), respectively. Lower panels show
       LDR-metallicity distribution for Co16757/Fe16661 (left panel), and Ti22310/Fe22257 (right panel).}
      \label{fig:ldr_mh}
\end{center}
\end{figure}

Metallicity dependence of LDRs has been previously discussed in 
several studies.
\cite{gray01} noted the effects of metallicity, absolute magnitude, and 
temperature on spectral lines. 
To improve the precision of the LDR-temperature calibrations in their work, 
they applied corrections for metallicity variations to the LDRs.
The \H-band studies of \cite{fukue15} and \cite{Jian19} also investigated
the effects of metallicity and abundance ratios on LDR$-$\teff\ relations. 
The \citeauthor{Jian19} analysis is especially useful, as they used
a large number of \H-band spectra (3700 $<$ \teff\ $<$ 5000K) from APOGEE 
survey with a metallicity range of $-$0.7 $<$ [Fe/H] $<$ 0.4 dex. 
They interpreted the saturations on abundance ratios to explain the 
metallicity effect on the LDRs.
Our sample has a metallicity range of $-$1.5 $<$ [M/H] $<$ 0.5. 
Investigating the metallicity range on LDRs, we found no clear relation 
between LDRs and the metallicities. 
An example plot that searches for the effect of [M/H] on LDR$-$\teff\ relations
is given in Figure~\ref{fig:ldr_mh}.
However, our stellar sample of 110 stars is too small to explore all
of the possible parameter dependencies of these newly-defined LDR
relationships.
Expansion of our sample size will be taken up in a future LDR calibration
study.

\subsection{Line-pairs with blended features}

For cool stars the near-infrared, especially the \K-band spectral 
region, is dominated by many molecular lines, making
it difficult to find unblended atomic lines. 
We noticed that some of the LDR$-$\teff\ 
calibrations listed in Table~\ref{tab-pairs} tend to give temperatures about 
100-200 K higher than the literature \teff\ values (Table~\ref{tab-sample}) 
when applied to stars that are not included in our sample (\S\ref{uncertain}): 
Arcturus, NGC 6940 MMU 105, NCG 752 MMU 77. These LDR$-$\teff\ calibrations 
are usually the ones that have atomic lines paired with \species{Sc}{i} 22052 
and 22065 \AA\ lines (also numbered as 15, 16, 17, 18, and 19 in Table~\ref{tab-pairs}). 
Closer examination of scandium lines indicated that \species{Sc}{i} 22052 line
blended with a \species{Si}{i} line (22051.93 \AA). 
Although the $\lambda$22065 \species{Sc}{i} line is unblended, 
a nearby CO feature becomes more intrusive at temperatures $\lesssim$ 4800 K.
Two additional line-pairs;  \species{Fe}{i} 20991/\species{Si}{i} 20926, and
\species{Ti}{i} 21897/\species{Fe}{i} 21895, were found to be also affected 
by the contribution of CO features that seem to become more potent especially 
for stars with \teff\ $\lesssim$ 4800 K. 
These LDR line pairs should be applied with caution.

\section{Summary and Conclusion}

In this study, we have applied the LDR method to high-resolution near infrared
spectra of 110 stars, of which were obtained with IGRINS in the \H\ and 
\K-band, simultaneously. To our knowledge, this is the first study that reports
the LDR$-$\teff\ relations in the \K-band spectral region.
Our sample covers mostly giant stars along with several
main-sequence stars and supergiants. Performing LDR$-$\teff\ calibrations
we found 26 LDR line-pairs that are mostly composed of Fe-peak elements 
(Table~\ref{tab-pairs}) and provide temperatures comparable
\textcolor{black}{ 
to literature \teff\ values within the accuracy of about 70 K. }
Investigation of our LDR$-$\teff\ relations
for the dependence of LDRs on \logg\ and [M/H] resulted in no clear relation.
We suggest further investigation of these parameters with more extensive sample 
drawn from the Raw and Reduced IGRINS Spectral Archive 
(RRISA\footnote{https://igrinscontact.github.io/}, \citealt{Sawczynec22}) in a future LDR study.
\\
\\

We thank our referee for helpful comments that improved this paper.
Our work has been supported by The Scientific and Technological
Research Council of Turkey (T\"{U}B\.{I}TAK, project No. 119F076), by the US
National Science Foundation grant AST-1616040 and by the
University of Texas Rex G. Baker, Jr. Centennial Research Endowment.
This work used the Immersion Grating Infrared Spectrometer (IGRINS) that 
was developed under a collaboration between the University of Texas at Austin 
and the Korea Astronomy and Space Science Institute (KASI) with the financial 
support of the Mt. Cuba Astronomical Foundation, of the US National Science 
Foundation under grants AST-1229522 and  AST-1702267, of the McDonald 
Observatory of the University of Texas at Austin, of the Korean GMT Project of 
KASI, and Gemini Observatory. This material is based upon work supported 
by the National Science Foundation under Grant No. AST-1908892 to G. Mace.
RLV acknowledges support from CONACYT through a postdoctoral fellowship 
within the program ``Estancias Posdoctorales por M\'exico".  
S{\"O} also acknowledges support by the National Science Centre, 
Poland, project 2019/34/E/ST9/00133.
This work has made use of data from the European Space Agency (ESA) mission
{\it Gaia} (\url{https://www.cosmos.esa.int/Gaia}), processed by the {\it Gaia}
Data Processing and Analysis Consortium (DPAC,
\url{https://www.cosmos.esa.int/web/Gaia/dpac/consortium}). Funding for the DPAC
has been provided by national institutions, in particular the institutions
participating in the {\it Gaia} Multilateral Agreement.
This publication makes use of data products from the Two Micron All Sky Survey, 
which is a joint project of the University of Massachusetts and the Infrared 
Processing and Analysis Center/California Institute of Technology, funded by 
the National Aeronautics and Space Administration and the National Science Foundation.
This research has made use of the SIMBAD database, operated at CDS, 
Strasbourg, France.

\software{IRAF (\citealt{tody93} and references therein),
          MOOG \citep{sneden73},
          ATLAS \citep{kurucz11}}

\clearpage

\bibliography{LDR}{}
\bibliographystyle{aasjournal}





\clearpage
\begin{center}
\begin{deluxetable}{lrrrrr}
\tabletypesize{\footnotesize}
\tablewidth{0pt}
\tablecaption{Atmospheric parameters of stellar sample.\label{tab-sample}}
\tablecolumns{6}
\tablehead{
\colhead{Star}                         &
\colhead{\teff\ (K)}                   &
\colhead{\logg}                        &
\colhead{[M/H]}                        &
\colhead{\vmicro\ (km/s)}              &
\colhead{Ref$\#$\tablenotemark{a}}                       
}
\startdata
HIP 80704	&	3250	&	0.20	&	0.07	&	2.70	&	1	\\
HIP 69038	&	3261	&	0.59	&	$-$0.02	&	2.00	&	2	\\
HIP 92791	&	3382	&	0.55	&	$-$0.06	&	2.30	&	3	\\
HIP 113881	&	3600	&	1.20	&	$-$0.11	&	2.00	&	1	\\
HIP 50801	&	3700	&	1.35	&	0.00	&	2.01	&	4	\\
HIP 67070	&	3700	&	1.00	&	$-$0.15	&	2.40	&	5	\\
HIP 77272	&	3935	&	0.59	&	$-$0.75	&	1.66	&	6	\\
HIP 26386	&	4000	&	1.25	&	$-$0.55	&	1.10	&	7	\\
HIP 102635	&	4010	&	1.78	&	$-$0.23	&	2.30	&	8	\\
HIP 3083	&	4048	&	1.13	&	0.10	&	2.10	&	3	\\
\enddata

\tablenotetext{a}{Reference Numbers:
(1)  \cite{smith85}, 
(2)  \cite{Koleva12}, 
(3)  \cite{Prugniel11}, 
(4)  \cite{Mallik98}, 
(5)  \cite{Smith90}, 
(6)  \cite{afsar18}, 
(7)  \cite{Pakhomov13}, 
(8)  \cite{McWilliam90}, 
(9)  \cite{Luck95}, 
(10) \cite{Luck14}, 
(11) \cite{jonsson17}, 
(12) \cite{Lyubimkov10}, 
(13) \cite{Feuillet16}, 
(14) \cite{Delgado17}, 
(15) \cite{Hekker07}, 
(16) \cite{ramirez13}, 
(17) \cite{Maldonado13}, 
(18) \cite{silva15}, 
(19) \cite{Jones11}, 
(20) \cite{jofre14}, 
(21) \cite{takeda08}, 
(22) \cite{Luck17}, 
(23) \cite{jofre15}, 
(24) \cite{Tabernero12}, 
(25) \cite{Takeda05},
(26) this study.
}

(This table is available in its entirety in machine-readable form.)

\end{deluxetable}
\end{center}

\clearpage
\begin{center}
\begin{deluxetable}{lccc}
\tabletypesize{\footnotesize}
\tablewidth{0pt}
\tablecaption{Atomic lines used for LDR pairs.\label{tab-lines}}
\tablecolumns{4}
\tablehead{
\colhead{Species}                   &
\colhead{Wavelength (\AA)}    &
\colhead{E.P. (eV)}          &
\colhead{log $gf$}              
}
\startdata
\species{V}{i}	&	15924.822	&	2.136	&	-1.200	\\
\species{Fe}{i}	&	15934.020	&	6.310	&	-0.430	\\
\species{Ti}{i}	&	16635.160	&	2.350	&	-1.582	\\
\species{Fe}{i}	&	16661.390	&	6.340	&	-0.070	\\
\species{Ni}{i}	&	16673.710	&	6.040	&	0.100	\\
\species{Co}{i}	&	16757.640	&	3.410	&	-1.310	\\
\species{Ti}{i}	&	17376.577	&	4.489	&	0.747	\\
\species{Fe}{i}	&	17420.880	&	3.882	&	-2.880	\\
\species{Fe}{i}	&	17433.635	&	6.411	&	0.030	\\
\species{Si}{i}	&	20917.151	&	6.727	&	0.575	\\
\species{Si}{i}	&	20926.149	&	6.727	&	-1.074	\\
\species{Fe}{i}	&	20991.037	&	4.143	&	-2.684	\\
\species{Si}{i}	&	21779.660	&	6.718	&	0.420	\\
\species{Ti}{i}	&	21782.940	&	1.749	&	-1.160	\\
\species{Si}{i}	&	21819.671	&	6.721	&	0.170	\\
\species{Si}{i}	&	21879.324	&	6.721	&	0.410	\\
\species{Fe}{i}	&	21894.996	&	6.144	&	-0.360	\\
\species{Ti}{i}	&	21897.370	&	1.739	&	-1.450	\\
\species{Ti}{i}	&	22004.500	&	1.730	&	-1.910	\\
\species{Sc}{i}	&	22051.985	&	1.448	&	-0.840	\\
\species{Na}{i}	&	22056.400	&	3.191	&	0.290	\\
\species{Si}{i}	&	22062.710	&	6.727	&	0.540	\\
\species{Sc}{i}	&	22065.306	&	1.439	&	-0.830	\\
\species{Fe}{i}	&	22079.852	&	5.849	&	-1.400	\\
\species{Na}{i}	&	22083.662	&	3.191	&	-0.013	\\
\species{Ti}{i}	&	22211.238		&	1.734	&	-1.780	\\
\species{Ti}{i}	&	22232.858	&	1.739	&	-1.690	\\
\species{Fe}{i}	&	22257.107	&	5.064	&	-0.710	\\
\species{Fe}{i}	&	22260.179	&	5.086	&	-0.941	\\
\species{Ti}{i}	&	22274.027	&	1.749	&	-1.790	\\
\species{Ti}{i}	&	22310.617	&	1.733	&	-2.071	\\
\species{Ti}{i}	&	22443.925	&	1.739	&	-2.370	\\
\enddata
\end{deluxetable}
\end{center}

\clearpage
\begin{center}
\begin{deluxetable}{rccccccrrr}
\tabletypesize{\footnotesize}
\tablewidth{0pt}
\tablecaption{LDR pairs and polynomial coefficients for LDR-\teff\ relations\tablenotemark{*}.\label{tab-pairs}}
\tablecolumns{10}
\tablehead{
\colhead{\#}                             &
\colhead{Species}                   &
\colhead{Wavelength (\AA)}    &
\colhead{Low$-$E.P. (eV)}                  &
\colhead{Species}                   &
\colhead{Wavelength (\AA)}    & 
\colhead{High$-$E.P. (eV)}      &
\colhead{a}   &
\colhead{b}   &
\colhead{c}   
}
\startdata
1	&	\species{V}{i}	&	15924.822	&	2.136	&	\species{Fe}{i}	&	15934.020	&	6.310	&	325.80	&	-1623.00	&	5302.2	\\
2	&	\species{Ti}{i}	&	16635.160	&	2.350	&	\species{Ni}{i}	&	16673.710	&	6.040	&	109.47	&	-940.02	&	5237.0	\\
3	&	\species{Co}{i}	&	16757.640	&	3.410	&	\species{Fe}{i}	&	16661.390	&	6.340	&	12.43	&	-1273.00	&	5440.5	\\
4	&	\species{Ti}{i}	&	17376.577	&	4.489	&	\species{Fe}{i}	&	17433.635	&	6.411	&	-544.70	&	-1615.90	&	5417.1	\\
5	&	\species{Fe}{i}	&	17420.880	&	3.882	&	\species{Fe}{i}	&	17433.635	&	6.411	&	172.50	&	-2039.50	&	5884.8	\\
6	&	\species{Fe}{i}	&	20991.037	&	4.143	&	\species{Si}{i}	&	20926.149	&	6.727	&	-57.92	&	-802.41	&	5521.3	\\
7	&	\species{Ti}{i}	&	21782.940	&	1.749	&	\species{Si}{i}	&	20917.151	&	6.727	&	-179.63	&	-570.72	&	5394.5	\\
8	&	\species{Ti}{i}	&	21782.940	&	1.749	&	\species{Si}{i}	&	21779.660	&	6.718	&	-24.00	&	-759.71	&	5501.4	\\
9	&	\species{Ti}{i}	&	21897.370	&	1.739	&	\species{Si}{i}	&	21779.660	&	6.718	&	21.27	&	-871.57	&	5382.1	\\
10	&	\species{Ti}{i}	&	21897.370	&	1.739	&	\species{Si}{i}	&	21819.671	&	6.721	&	-58.17	&	-661.10	&	5360.5	\\
11	&	\species{Ti}{i}	&	21897.370	&	1.739	&	\species{Si}{i}	&	21879.324	&	6.721	&	-45.67	&	-714.48	&	5340.6	\\
12	&	\species{Ti}{i}	&	21897.370	&	1.739	&	\species{Fe}{i}	&	21894.996	&	6.144	&	-239.59	&	-155.53	&	5341.8	\\
13	&	\species{Ti}{i}	&	22004.500	&	1.730	&	\species{Na}{i}	&	22056.400	&	3.191	&	-1311.80	&	-442.95	&	5193.8	\\
14	&	\species{Ti}{i}	&	22004.500	&	1.730	&	\species{Si}{i}	&	22062.710	&	6.727	&	71.95	&	-991.17	&	5274.7	\\
15	&	\species{Sc}{i}	&	22051.985	&	1.448	&	\species{Si}{i}	&	21779.660	&	6.718	&	117.55	&	-966.01	&	5208.8	\\
16	&	\species{Sc}{i}	&	22051.985	&	1.448	&	\species{Si}{i}	&	22062.710	&	6.727	&	120.91	&	-962.31	&	5204.5	\\
17	&	\species{Sc}{i}	&	22065.306	&	1.439	&	\species{Si}{i}	&	22062.710	&	6.727	&	158.02	&	-1098.40	&	5135.6	\\
18	&	\species{Sc}{i}	&	22065.306	&	1.439	&	\species{Fe}{i}	&	22079.852	&	5.849	&	4.70	        &	-230.17	&	5205.1	\\
19	&	\species{Sc}{i}	&	22065.306	&	1.439	&	\species{Na}{i}	&	22083.662	&	3.191	&	429.44	&	-1988.20	&	5182.1	\\
20	&	\species{Ti}{i}	&	22211.238		&	1.734	&	\species{Fe}{i}	&	22257.107	&	5.064	&	-285.66	&	-581.50	&	5260.3	\\
21	&	\species{Ti}{i}	&	22211.238		&	1.734	&	\species{Fe}{i}	&	22260.179	&	5.086	&	-164.53	&	-694.44	&	5323.4	\\
22	&	\species{Ti}{i}	&	22232.858	&	1.739	&	\species{Fe}{i}	&	22257.107	&	5.064	&	-341.30	&	-439.41	&	5248.7	\\
23	&	\species{Ti}{i}	&	22232.858	&	1.739	&	\species{Fe}{i}	&	22260.179	&	5.086	&	-135.50	&	-707.92	&	5374.1	\\
24	&	\species{Ti}{i}	&	22274.027	&	1.749	&	\species{Fe}{i}	&	22257.107	&	5.064	&	-405.20	&	-428.16	&	5225.9	\\
25	&	\species{Ti}{i}	&	22310.617	&	1.733	&	\species{Fe}{i}	&	22257.107	&	5.064	&	-42.72	&	-1135.70	&	5262.3	\\
26	&	\species{Ti}{i}	&	22443.925	&	1.739	&	\species{Fe}{i}	&	22260.179	&	5.086	&	219.42	&	-1518.80	&	5259.2	\\
\enddata

\tablenotetext{*}{\textcolor{black}{ All LDR-\teff\ relations are represented with quadratic polynomials (\teff=a$x^{2}$+b$x$+c), 
in which the $x$ is the LDR value.}}

\end{deluxetable}
\end{center}

\clearpage

\begin{center}
\begin{deluxetable}{lrrrr}
\tabletypesize{\footnotesize}
\tablewidth{0pt}
\tablecaption{\teff\ values of our sample from various sources: spectral \teff\ values 
                     (Table~\ref{tab-sample}), $\overline{\tldr}$ (this study), \teff({\it Gaia}), and \teff(J-K).\label{tab-comp}}
\tablecolumns{5}
\tablehead{
\colhead{Star}                         &
\colhead{\teff\ (K)}                   &
\colhead{\tldr}                        &
\colhead{\teff({\it Gaia})}                        &
\colhead{\teff(J$-$K)}                      
}
\startdata
HIP 80704 	&	3250	&	3279	&		&	3555	\\
HIP 69038 	&	3261	&	3383	&		&	3451	\\
HIP 92791 	&	3382	&	3408	&		&	3610	\\
HIP 113881	&	3600	&	3480	&		&	3982	\\
HIP 67070 	&	3700	&	3694	&	3851	&	3711	\\
HIP 50801 	&	3700	&	3705	&		&	4084	\\
HIP 77272 	&	3935	&	4008	&	4065	&	3727	\\
HIP 26386 	&	4000	&	3866	&	4250	&	3695	\\
HIP 102635	&	4010	&	4006	&	4061	&	4014	\\
HIP 3083  	&	4048	&	3939	&	4017	&	3802	\\
\enddata

(This table is available in its entirety in machine-readable form.)

\end{deluxetable}
\end{center}

\clearpage

\end{document}